# Double version of the Rashba and Dresselhaus spin-orbit coupling

Hu Zhang*, Lulu Zhao, Chendong Jin, Ruqian Lian, Peng-Lai Gong, RuiNing Wang, JiangLong Wang, and Xing-Qiang Shi

Key Laboratory of Optic-Electronic Information and Materials of Hebei Province, Institute of Life Science and Green Development, College of Physics Science and Technology, Hebei University, Baoding 071002, P. R. China

* E-mails: zhanghu@hbu.edu.cn

The Rashba and Dresselhaus types of spin-orbit coupling are two typical linear coupling forms. We establish the fundamental physics of a model which can be viewed as the double version of the Rashba and Dresselhaus spin-orbit coupling. This model describes the low energy physics of a class of massless Dirac fermions in spin-orbit systems. The physical properties of the massless Dirac fermions are determined by the mathematical relations of spin-orbit coefficients. For equal Rashba and Dresselhauss coupling constants, $k$-independent eigenspinors and a persistent spin helix combined with massless birefringent Dirac fermions emerge in this model. The spin-orbit coupled systems described by this model have potential technological applications from spintronics to quantum computation.

The spin-orbit coupling described by the Dirac equation and its nonrelativistic expansion is important to understand the physics of semiconductor spintronics, quantum computing, the spin Hall effect and topological phase of matters [1-6]. One of the typical linear coupling forms is the Rashba spin-orbit coupling:

$$H_{\rm R} = \alpha_{\rm R}(k_y\sigma_x - k_x\sigma_y), \tag{1}$$

where $\alpha_{\rm R}$ is the Rashba spin-orbit coefficient, $k_{x,y}$ is the electron wave vectors, and $\sigma_{x,y}$ are the Pauli matrices [7,8]. Based on the Rashba spin-orbit coupling, Datta and Das proposed the concept of spin field effect transistor (spin FET) [9]. Another linear coupling form is known as the Dresselhauss spin-orbit coupling given by:

$$H_\text{D} = \beta_\text{D}(k_x\sigma_x - k_y\sigma_y), \tag{2}$$

where $\beta_\text{D}$ is the Dresselhauss spin-orbit coefficient [10]. This Hamiltonian describes the spin-orbit coupling in a two-dimensional semiconductor thin film grown with appropriate geometry.

In a two-dimensional spin-orbit coupled system without inversion symmetry, the coexistence of Rashba and Dresselhaus contributions is possible [11]. The Hamiltonian is of the form:

$$H_\text{RD} = \frac{\hbar^2 k^2}{2m} + H_\text{R} + H_\text{D}, \tag{3}$$

where $m$ is the effective electron mass [12]. Due to the tunable of the Rashba term by gating, the Rashba term and the Dresselhaus term can have equal strengths $\alpha_\text{R} = \beta_\text{D}$. This situation leads to $k$-independent eigenspinors in two dimensions. A nonballistic spin FET was proposed based on this unique feature [11]. Furthermore, a persistent spin helix was predicted for the special case of $\alpha_\text{R} = \beta_\text{D}$ [2,12]. In 2009, the emergence of the persistent spin helix in GaAs quantum wells was observed experimentally [13].

In this work, we consider a model described by a double version of the Rashba and Dresselhaus spin-orbit coupling. The Hamiltonian is given by:

$$H_\text{Double} = \begin{pmatrix} \alpha_1(k_y\sigma_x - k_x\sigma_y) + \beta_1(k_x\sigma_x - k_y\sigma_y) & 0 \\ 0 & \alpha_2(k_y\sigma_x - k_x\sigma_y) + \beta_2(k_x\sigma_x - k_y\sigma_y) \end{pmatrix}, \tag{4}$$

where $\alpha_{1,2}$ and $\beta_{1,2}$ are the strengths of the Rashba and Dresselhauss spin-orbit couplings respectively. The four four-component eigenstates of the Hamiltonian $H_\text{Double}$ are:

$$\varphi^{(1)} = \frac{1}{\sqrt{2}}\begin{pmatrix} -e^{i\theta_1(\vec{k})} \\ 1 \\ 0 \\ 0 \end{pmatrix}, \quad \varphi^{(2)} = \frac{1}{\sqrt{2}}\begin{pmatrix} e^{i\theta_1(\vec{k})} \\ 1 \\ 0 \\ 0 \end{pmatrix}, \quad \varphi^{(3)} = \frac{1}{\sqrt{2}}\begin{pmatrix} 0 \\ 0 \\ -e^{i\theta_2(\vec{k})} \\ 1 \end{pmatrix}, \quad \varphi^{(4)} = \frac{1}{\sqrt{2}}\begin{pmatrix} 0 \\ 0 \\ e^{i\theta_2(\vec{k})} \\ 1 \end{pmatrix}, \tag{5}$$

with $\theta_1(\vec{k}) = \arg[\alpha_1 k_y + \beta_1 k_x + i(\alpha_1 k_x + \beta_1 k_y)]$, $\theta_2(\vec{k}) = \arg[\alpha_2 k_y + \beta_2 k_x + i(\alpha_2 k_x + \beta_2 k_y)]$ and with the corresponding four eigenenergies:

$$\varepsilon^{(1)}(\vec{k}) = -\sqrt{(\alpha_1 k_y + \beta_1 k_x)^2 + (\alpha_1 k_x + \beta_1 k_y)^2},$$

$$\varepsilon^{(2)}(\vec{k}) = \sqrt{(\alpha_1 k_y + \beta_1 k_x)^2 + (\alpha_1 k_x + \beta_1 k_y)^2},$$

$$\varepsilon^{(3)}(\vec{k}) = -\sqrt{(\alpha_2 k_y + \beta_2 k_x)^2 + (\alpha_2 k_x + \beta_2 k_y)^2},$$

$$\varepsilon^{(4)}(\vec{k}) = \sqrt{(\alpha_2 k_y + \beta_2 k_x)^2 + (\alpha_2 k_x + \beta_2 k_y)^2}. \tag{6}$$

Therefore, we have two positive energy solutions (particles) and two negative energy solutions (holes) for this model. The Fermi surfaces of the two positive energy solutions $\varepsilon^{(2)}$ and $\varepsilon^{(4)}$ at the Fermi level $E_F$ are, respectively, determined by:

$$(\alpha_1 k_y + \beta_1 k_x)^2 + (\alpha_1 k_x + \beta_1 k_y)^2 = E_F^2,$$

$$(\alpha_2 k_y + \beta_2 k_x)^2 + (\alpha_2 k_x + \beta_2 k_y)^2 = E_F^2. \tag{7}$$

The Fermi surfaces of the two negative energy solutions have the same form. At the $k_x = k_y = 0$ point, the four eigenenergies are all zero and thus the four solutions are degenerate. This four-fold degenerate point in the momentum space is called the Dirac point [14]. In fact, the Hamiltonian $H_{\text{Double}}$ describes the low energy physics of a class of *massless* Dirac fermions in spin-orbit systems. Hence $k_{x,y}$ can be viewed as the electron wave vectors related to the Dirac point. In experiment, the Rashba and Dresselhaus spin-orbit coupling can be realized in cold atoms [15-18]. Thus, the model in (4) might be realized with cold atoms in an optical lattice. The significant differences between these massless Dirac fermions and ones previously identified in Dirac semimetals [14] will be discussed below. For general $\alpha_{1,2}$ and $\beta_{1,2}$, the spinor in the eigenstates (5) depends on the wave vector via $\theta_{1,2}(\vec{k})$ and the Fermi velocity of the massless Dirac fermions is $k$-dependent. For spin-orbit coefficients $\alpha_{1,2}$ and $\beta_{1,2}$ with different mathematical relations, the physical properties of the massless Dirac fermions are significantly different. These will be discussed case by case in the following.

Firstly, we consider the case of $\beta_1 = 0 = \beta_2$, which reflects the absence of the Dresselhaus spin-orbit coupling in Hamiltonian $H_{\text{Double}}$. There are still two mathematical relations $\alpha_1 = \alpha_2$ and $\alpha_1 \neq \alpha_2$ for the Rashba spin-orbit coupling coefficients $\alpha_{1,2}$. For $\alpha_1 = \alpha_2$, the energy bands have the relations $\varepsilon^{(1)}(\vec{k}) = \varepsilon^{(3)}(\vec{k})$ and $\varepsilon^{(2)}(\vec{k}) = \varepsilon^{(4)}(\vec{k})$. Thus, the two positive energy solutions and two negative energy solutions are degenerate respectively. Such band dispersion relation gives two overlapped Dirac cones in the $k_x$-$k_y$ plane. We have plotted the Dirac cones in Fig. 1(a). The Hamiltonian $H_{\text{Double}}$ describes the low energy physics of the massless Dirac fermions with the Fermi velocity of $\alpha_1$. From (7) we find that the Fermi surfaces consist of two overlapped circles with radius of $|E_F/\alpha_1|$, which are shown in Fig. 1(b). In this case, $\theta_1(\vec{k})$ and $\theta_2(\vec{k})$ are independent of $\alpha_1$ and $\alpha_2$ respectively. For $\alpha_1 \neq \alpha_2$, the degenerated energy bands split. We have plotted the Dirac cones in Fig. 1(c). Now the massless Dirac fermions have two different Fermi velocity of $\alpha_1$ and $\alpha_2$, which can be called the massless *birefringent* Dirac fermions [19] to exhibit the split nature of the doubly degenerate Dirac cones. The Fermi surfaces consist of two circles with radii of $|E_F/\alpha_{1,2}|$ as shown in Fig. 1(d). In fact, this model descries the low energy physics of massless birefringent Dirac fermions in some three-dimensional Dirac semimetals with a polar crystal structure [20].

In the Hamiltonian $H_{\text{Double}}$, we can also let $\alpha_1 = 0 = \alpha_2$. This situation corresponds to the absence of the Rashba spin-orbit coupling and thus only the Dresselhaus term remains. For $\beta_1 = \beta_2$, the energy bands have the relations $\varepsilon^{(1)}(\vec{k}) = \varepsilon^{(3)}(\vec{k})$ and $\varepsilon^{(2)}(\vec{k}) = \varepsilon^{(4)}(\vec{k})$. The Hamiltonian $H_{\text{Double}}$ describes the low energy physics of the massless Dirac fermions with the Fermi velocity of $\beta_1$. The Dirac cones are similar to the ones in Fig. 1(a) but with different slopes. Now, $\theta_1(\vec{k})$ and $\theta_2(\vec{k})$ are independent of $\beta_1$ and $\beta_2$ respectively. It should be noted that $k_x$ and $k_y$ in $\theta_{1,2}(\vec{k})$ are interchanged compared to those for $\beta_1 = 0 = \beta_2$. For $\beta_1 \neq \beta_2$, the degenerated energy bands split. This results in massless birefringent Dirac fermions with the Fermi velocity of $\beta_1$ and $\beta_2$. The Dirac cones and the corresponding Fermi surfaces are similar to those in Fig. 1(c, d). For both $\beta_1 \neq \beta_2$ and $\beta_1 = \beta_2$, $\theta_{1,2}(\vec{k})$ have the same form.

Now we consider the case of both $\alpha_{1,2} \neq 0$ and $\beta_{1,2} \neq 0$, which corresponds to the coexistence of the Rashba and Dresselhaus spin-orbit coupling in $H_{\text{Double}}$. For $\alpha_1 \neq \pm\beta_1$ and $\alpha_2 \neq \pm\beta_2$, the equations of the Fermi surfaces given in (7) describe two ellipses and can be rewritten as:

$$k_x^2 + k_y^2 + \gamma_1 k_x k_y = \frac{E_F^2}{\alpha_1^2 + \beta_1^2},$$
$$k_x^2 + k_y^2 + \gamma_2 k_x k_y = \frac{E_F^2}{\alpha_2^2 + \beta_2^2}, \quad (8)$$

where $\gamma_1 = \frac{4\alpha_1\beta_1}{\alpha_1^2+\beta_1^2}$ and $\gamma_2 = \frac{4\alpha_2\beta_2}{\alpha_2^2+\beta_2^2}$. Rotating the spatial coordinates by introducing $k_\pm = (1/\sqrt{2})(k_x \pm k_y)$ brings (8) to the form:

$$\frac{k_+^2}{\zeta_1^2} + \frac{k_-^2}{\eta_1^2} = 1,$$
$$\frac{k_+^2}{\zeta_2^2} + \frac{k_-^2}{\eta_2^2} = 1, \quad (9)$$

with $\zeta_{1,2}^2 = \frac{E_F^2}{(\alpha_{1,2}+\beta_{1,2})^2}$ and $\eta_{1,2}^2 = \frac{E_F^2}{(\alpha_{1,2}-\beta_{1,2})^2}$. These equations describe two ellipses with the standard form lying in the rotated spatial coordinate axes $k_\pm$. From (7) or (9) one can also find out the tangency points of the two ellipses for particular $\alpha_{1,2}$ and $\beta_{1,2}$. When $(\alpha_1 + \beta_1)^2 = (\alpha_2 + \beta_2)^2$ the tangency points satisfy the relation $k_x - k_y = 0$. For $(\alpha_1 - \beta_1)^2 = (\alpha_2 - \beta_2)^2$, the tangency points satisfy the relation $k_x + k_y = 0$. The Dirac cones and the corresponding Fermi surfaces for the latter case were plotted in Fig. 2(a, b). The Fermi velocities of this type of Dirac fermions are $k$-dependent exhibiting strong anisotropy. For electrons with wave vectors satisfying $k_x + k_y = 0$, the Fermi velocity has only one value of $\sqrt{2}|\alpha_1 - \beta_1|$. At $k_x = 0$ (or $k_y = 0$), the Fermi velocities are $\sqrt{\alpha_1^2 + \beta_1^2}$ and $\sqrt{\alpha_2^2 + \beta_2^2}$. Now, $\theta_1(\vec{k})$ and $\theta_2(\vec{k})$ depend on wave vectors as well as spin-orbit coefficients. On the other hand, from the general properties of ellipses we have the following results: the two ellipses in (9) have no crossing points when $(\alpha_1 + \beta_1)^2 > (\alpha_2 + \beta_2)^2$ and $(\alpha_1 - \beta_1)^2 > (\alpha_2 - \beta_2)^2$ (or $(\alpha_1 + \beta_1)^2 < (\alpha_2 + \beta_2)^2$ and $(\alpha_1 - \beta_1)^2 < (\alpha_2 - \beta_2)^2$); the two ellipses have four crossing points when $(\alpha_1 + \beta_1)^2 > (\alpha_2 + \beta_2)^2$ and $(\alpha_1 - \beta_1)^2 < (\alpha_2 - \beta_2)^2$ (or $(\alpha_1 + \beta_1)^2 < (\alpha_2 + \beta_2)^2$ and $(\alpha_1 - \beta_1)^2 > (\alpha_2 - \beta_2)^2$). In Fig. 2(c, d), we have plotted the crossed Dirac cones and the corresponding Fermi surfaces having four crossing points. Again, both the Fermi velocities and $\theta_{1,2}(\vec{k})$ of this type of Dirac fermions show strong anisotropy.

For $\alpha_1 = \beta_1$ and $\alpha_2 = \beta_2$ ($\alpha_1 \neq \alpha_2$), the Fermi surfaces in (7) reduce to $(k_x + k_y)^2 = E_F^2/2\alpha_{1,2}^2$, which is the equation of four parallel lines. From (6) we know that wo positive energy solutions and two negative energy solutions are all zero and thus degenerate at the line $k_x + k_y = 0$. We have plotted the band dispersions and corresponding Fermi surfaces in Fig. 3(a, b). Evidently, the Dirac cones become crossing planes looking like a butterfly. The Fermi velocities of such massless birefringent Dirac fermions are $\sqrt{2}\alpha_{1,2}|k_x + k_y|/\sqrt{k_x^2 + k_y^2}$, which also show strong anisotropy. These results indicate that the line $k_x + k_y = 0$ behaves as a nodal line of this system. From the mathematical relations $\alpha_1 = \beta_1$ and $\alpha_2 = \beta_2$ we obtain $\theta_1(\vec{k}) = \theta_2(\vec{k}) = \pi/4$, which means that four-component eigenstates in (6) are $k$-independent. This property is related to the fact that the spin operator $\Sigma = (1/\sqrt{2})\sigma_0 \otimes (\sigma_x - \sigma_y)$, in which $\sigma_0$ is the identity matrix, is a conserved quantity in the current case,

$$[H_{\text{Double}}, \Sigma] = 0. \qquad (10)$$

Due to the absence of the linear term in $H_{\text{Double}}$ at the line $k_x + k_y = 0$, we may consider high order term in $k$. Motivated by (3), we add the diagonal term $\hbar^2 k^2/2m$ into the Hamiltonian $H_{\text{Double}}$, which breaks the symmetry between particles and holes. Now the Dirac fermions along the line $k_x + k_y = 0$ have parabolic dispersions. The cones in this case were plotted in Fig. 3(c). The corresponding Fermi surfaces shown in Fig. 3(d) exist singular points at $k_x = -k_y$. Since the $k^2$ term is spin independent, the spin operator $\Sigma$ is still a conserved quantity. We still have $\theta_1(\vec{k}) = \theta_2(\vec{k}) = \pi/4$ and thus the eigenstates are also $k$-independent. If $\alpha_1 = \alpha_2$, the two positive energy solutions and two negative energy solutions are degenerate respectively.

The existence of the $k$-independent eigenstates allows the design of the nonballistic spin FET based on such massless birefringent Dirac fermions in spin-orbit coupling systems (as a channel) described by $H_{\text{Double}}$. Similar to the case discussed by Schliemann *et. al.* based on the model $H_{\text{RD}}$ given by (3), the "ON" and "OFF" states of our transistor based on $H_{\text{Double}}$ correspond to a gate bias for which $\alpha_1 = \beta_1$, $\alpha_2 = \beta_2$ and $\alpha_1 \neq \beta_1$, $\alpha_2 \neq \beta_2$ respectively. For the "ON" state, the injected electron is in one of the *k*-independent spin states ($\pm \exp(i\pi/4)$, 1) and traverses the channel (with $E_F > 0$) without changing its spin state. For the "OFF" state, the injected spin state of the electron is *k*-dependent and will be randomized when it traverses the channel due to strong Dyakonov-Perel type spin relaxation and results in unpolarized output current.

Finally, we show that the condition $\alpha_1 = \beta_1$ and $\alpha_2 = \beta_2$ ($\alpha_1 \neq \alpha_2$) leads to the existence of the persistent spin helix resulting from the precession of the diffusing spin under the spin-orbit effective magnetic field. The effect of the spin-orbit effective magnetic field is the spilt of the energy band $\varepsilon^{(2)}$ and $\varepsilon^{(4)}$ with the value of $\sqrt{2}(|\alpha_1| - |\alpha_2|)(k_x + k_y)$ with or without considering the $k^2$ term. For an electron traveling in the plane with wave vector $\vec{k}$, its spin precesses about the $x_-$ axis (related to the spin operator Σ defined above) by an angle

$$\phi = \frac{\sqrt{2}(|\alpha_1| - |\alpha_2|)(k_x + k_y)t}{\hbar} = \frac{2(|\alpha_1| - |\alpha_2|)k_+ t}{\hbar}, \tag{11}$$

in time *t*. The distance of the electron along the $x_+$ direction is $L_+ = \hbar k_+ t/m$. Then we have $\phi = 2(|\alpha_1| - |\alpha_2|)mL_+/\hbar^2$. This means that the net spin precession in the *x*, *y* plane depends only on the net displacement of an electron in the $x_+$ direction and is independent of any other property of its path. As shown in Fig. 4, electrons starting with parallel spin orientation and the same value of $x_+$ will return exactly to the original orientation ($\phi = 2\pi$) after propagating $L_+ = \pi\hbar^2/(|\alpha_1| - |\alpha_2|)m$. This is the physical origin of the persistent spin helix based on the Hamiltonian $H_{\text{Double}}$.

In summary, we have proposed a model which is a double version of the Rashba and Dresselhaus spin-orbit coupling. This model describes the low energy physics of a class of massless Dirac fermions in spin-orbit systems. The different mathematical relations of spin-orbit coefficients lead significantly different physical properties for the massless Dirac fermions. The various Fermi surfaces consisting of circles, ellipses, lines, and the ones with singular points are found. We also predict the existence of *k*-independent eigenspinors and the persistent spin helix for models with equal Rashba and Dresselhauss spin-orbit coupling constants. Based on the physical properties of massless Dirac fermions, these spin-orbit coupled systems have potential technological applications such as spintronics.

This work was supported by the Advanced Talents Incubation Program of the Hebei University (Grants No. 521000981423, No. 521000981394, No. 521000981395, and No. 521000981390), the Natural Science Foundation of Hebei Province of China (Grants No. A2021201001 and No. A2021201008), the National Natural Science Foundation of China (Grants No. 12104124, No. 11904154, and No. 51772297), and the high-performance computing center of Hebei University.

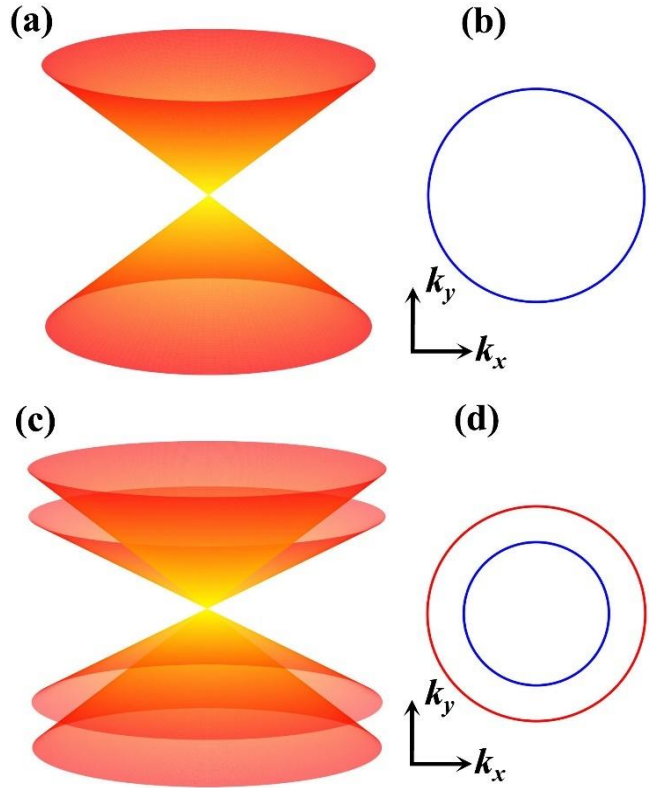

FIG. 1. (a) The band dispersion in the $k_x$-$k_y$ plane and (b) the corresponding Fermi surface for $\beta_1 = 0 = \beta_2$ and $\alpha_1 = \alpha_2$. (c) The band dispersion in the $k_x$-$k_y$ plane and (d) the corresponding Fermi surface for $\beta_1 = 0 = \beta_2$ and $\alpha_1 \neq \alpha_2$.

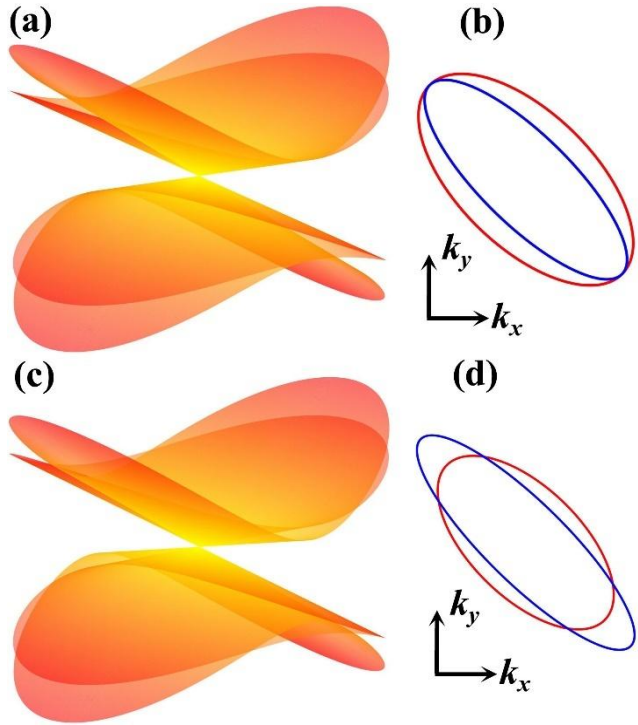

FIG. 2. (a) The band dispersion in the $k_x$-$k_y$ plane and (b) the corresponding Fermi surface for

$(\alpha_1 - \beta_1)^2 = (\alpha_2 - \beta_2)^2$ ($\alpha_{1,2} \neq 0$ and $\beta_{1,2} \neq 0$). (c) The band dispersion in the $k_x$-$k_y$ plane and (d) the corresponding Fermi surface for $(\alpha_1 + \beta_1)^2 < (\alpha_2 + \beta_2)^2$ and $(\alpha_1 - \beta_1)^2 > (\alpha_2 - \beta_2)^2$ ($\alpha_{1,2} \neq 0$ and $\beta_{1,2} \neq 0$).

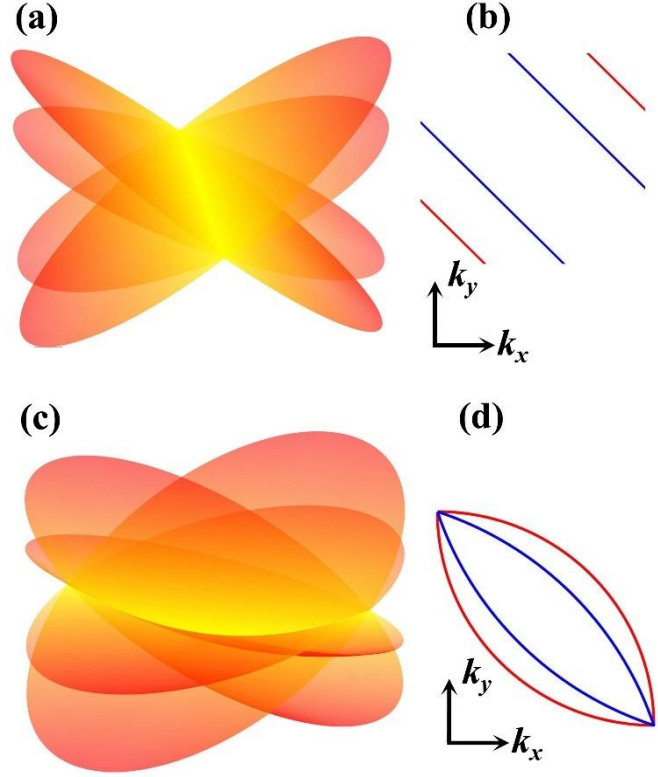

FIG. 3. (a) The band dispersion in the $k_x$-$k_y$ plane and (b) the corresponding Fermi surface for $\alpha_1 = \beta_1$ and $\alpha_2 = \beta_2$ ($\alpha_1 \neq \alpha_2$). (c) The band dispersion in the $k_x$-$k_y$ plane and (d) the corresponding Fermi surface for $\alpha_1 = \beta_1$ and $\alpha_2 = \beta_2$ ($\alpha_1 \neq \alpha_2$) with the consideration of the term $\hbar^2 k^2 / 2m$ in the Hamiltonian.

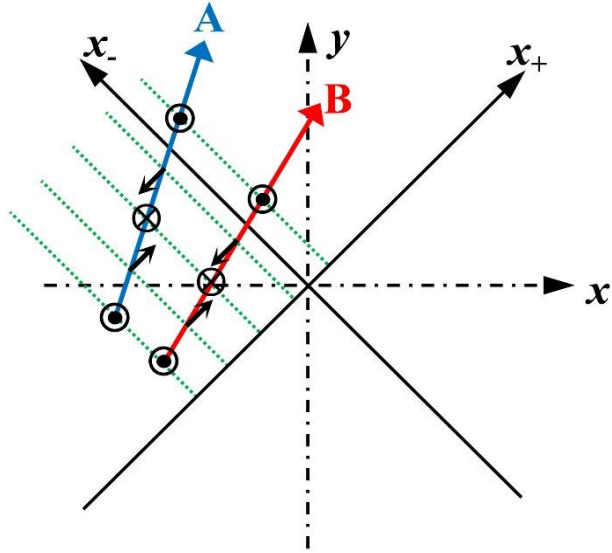

FIG. 4. Spin configurations for electrons A and B depend only on their distance traveled along the $x_+$ axis. The spins all return to the original configuration after propagating $L_+ = \pi\hbar^2/(|\alpha_1| - |\alpha_2|)m$.